\documentclass[12pt]{article}
\usepackage{times}
\usepackage{geometry}
\usepackage{graphicx}
\usepackage[dvipsnames]{xcolor}
\usepackage{natbib}
\usepackage{enumitem,amssymb}
\usepackage{ragged2e}
\newlist{thematic}{itemize}{8}
\setlist[thematic]{label=$\square$}
\usepackage{pifont}

\usepackage{hyperref}
\newcommand\myshade{100}
\colorlet{mylinkcolor}{violet}
\colorlet{mycitecolor}{YellowOrange}
\colorlet{myurlcolor}{Aquamarine}
\hypersetup{
  linkcolor  = mylinkcolor!\myshade!black,
  citecolor  = mycitecolor!\myshade!black,
  urlcolor   = myurlcolor!\myshade!black,
  colorlinks = true,
}
\usepackage[capitalize]{cleveref}

\DeclareRobustCommand{\ion}[2]{\textup{#1\,\textsc{\lowercase{#2}}}}

\begin{document}

\title{International Astrophysical Consortium \\for High-energy Calibration
\\
\bf{Summary of the 18th IACHEC Workshop}
}

\author{
C.~E.~Grant$^1$,
K.~K.~Madsen$^2$,
V.~Burwitz$^3$,
H. Earnshaw$^{4}$,
K.~Forster$^4$,
M.~Guainazzi$^5$, \and
V.L.~Kashyap$^6$, 
H.~L.~Marshall$^1$,
C.~B. Markwardt$^2$,
F.~Meeker-F\"urst$^{11}$,
E.~D.~Miller$^1$,  \and
L.~Natalucci$^7$, 
P.~P.~Plucinsky$^6$,
M.~Shidatsu$^{8}$,
H.~Suzuki$^{12}$,
Y.~Terada$^{9,10}$
A. T\"{u}mer$^{13,2}$
}

\date{\today}   

\newcommand{\artxc}{\textit{ART-XC}}
\newcommand{\erosita}{\textit{eROSITA}}
\newcommand{\rosat}{\textit{ROSAT}}
\newcommand{\xmm}{XMM-\textit{Newton}}
\newcommand{\chandra}{\textit{Chandra}}
\newcommand{\suzaku}{\textit{Suzaku}}
\newcommand{\swift}{\textit{Swift}}
\newcommand{\nicer}{\textit{NICER}}
\newcommand{\astrosat}{\textit{AstroSat}}
\newcommand{\nustar}{\textit{NuSTAR}}
\newcommand{\hxmt}{\textit{Insight-HXMT}}
\newcommand{\hitomi}{\textit{Hitomi}}
\newcommand{\xrism}{\textit{XRISM}}
\newcommand{\integral}{\textit{INTEGRAL}}
\newcommand{\fermi}{\textit{Fermi}}
\newcommand{\athena}{\textit{Athena}}
\newcommand{\newathena}{\textit{NewAthena}}
\newcommand{\ep}{\textit{Einstein Probe}}
\newcommand{\ixpe}{\textit{IXPE}}
\newcommand{\leia}{\textit{LEIA}}
\newcommand{\rxte}{\textit{RXTE}}
\newcommand{\EP}{\textit{Einstein Probe}}
\DeclareRobustCommand{\ion}[2]{\textup{#1\,\textsc{\lowercase{#2}}}}
\newcommand{\cstat}{{\tt c-stat}}
\newcommand{\etal}{\mbox{{\em et~al.}}}

\maketitle

{\centering
  $^1$Kavli Institute for Astrophysics and Space Research, \\ Massachusetts Institute of Technology, USA \\  
  $^2$NASA Goddard Space Flight Center, USA \\
  $^3$Max Planck Institute for Extraterrestrial Physics, Germany\\
  $^4$Cahill Center for Astronomy and Astrophysics, California Institute of Technology, USA \\
  $^5$ESA-ESTEC, The Netherlands \\
  $^6$Center for Astrophysics $|$ Harvard \& Smithsonian (CfA), USA \\
  $^7$IAPS-INAF, Italy \\
  $^8$Ehime University, Department of Physics, Japan \\
  $^9$Saitama University, Graduate School of Science and Engineering, Japan \\
  $^{10}$Japan Aerospace Exploration Agency, Institute of Space and Astronautical Science, Japan \\
  $^{11}$European Space Agency (ESA), European Space Astronomy Center (ESAC), Camino Bajo del Castillo s/n, 28692 {Villanueva de la Ca\~nada, Madrid}, {Spain} \\
  $^{12}$University of Miyazaki, Faculty of Engineering, Japan \\
$^{13}$Center for Space Science and Technology, University of Maryland, Baltimore County (UMBC), Baltimore, MD 21250, USA \\
}
\vspace{5mm}
\begin{center}
{\bf \large Abstract} \\
\end{center}

In this report we summarize the activities of the International Astronomical Consortium for High Energy Calibration (IACHEC) from the 18th IACHEC Workshop at Pelham, Germany. Fifty-six scientists directly involved in the calibration of operational and future high-energy missions gathered during four days to discuss the status of the cross-calibration between the current international complement of X-ray observatories, and the possibilities to improve it. This summary consists of reports from the Working Groups with topics ranging across: the identification and characterization of standard calibration sources, multi-observatory cross-calibration campaigns, appropriate and new statistical techniques, calibration of instruments and characterization of background, preservation of knowledge, and results for the benefit of the astronomical community.

\section{Meeting Summary}

The International Astronomical Consortium for High Energy Calibration (IACHEC)\footnote{\href{http://iachec.org}{\tt http://iachec.org}} is a group dedicated to supporting the cross-calibration of the scientific payload of high energy astrophysics missions with the goal of maximizing their scientific return. Its members are drawn from instrument teams, international and national space agencies, and other scientists with an interest in calibration. Representatives of over a dozen current and future missions regularly contribute to the IACHEC activities. 

The 18th International Astronomical Consortium for High Energy Calibration (IACHEC) workshop took place in Germany in 2026, attracting fifty-six participants with over sixty presentations. This meeting represented a continuation of the IACHEC tradition that began in 2006 in Iceland, with the consortium having grown significantly over two decades. The meeting featured a diverse representation from X-ray astronomy missions spanning multiple generations, including the veteran Chandra and XMM-Newton observatories, the recently launched XRISM mission, the Indian AstroSAT that achieved its tenth anniversary milestone in September 2025, and newer entrants such as Einstein Probe, SVOM, and XPoSat, with upcoming missions like SMILE and COSI planned for 2026 and 2027 respectively. The consortium also faced sobering realities as Swift required an emergency intervention that unfortunately didn't succeed, and NICER operations were discontinued, underscoring the challenges of maintaining the aging fleet of X-ray observatories.

The organizational structure of the 2026 IACHEC workshop emphasized the evolving purpose of these gatherings. While working groups historically prioritized in-meeting collaborative work to develop standards, cross-calibrate instruments, and curate data, the meeting format has shifted toward providing a forum where missions share their calibration challenges, learn from each other's experiences, and establish crucial interpersonal connections within the community. The 2026 meeting featured two new working groups: the Mission and Science Operations Working Group chaired by Felix Fürst from ESA with Hannah Earnshaw from Caltech as co-chair, which presented four talks from XMM, NICER, NuSTAR, and XRISM; and the Ground Calibration Working Group led by Vadim Burwitz from MPE and Takashi Okajima from Goddard Space Flight Center, which included two talks, a paper proposal, and discussions about sharing reference optics, defining analysis standards, and incorporating absolute calibration standards.

The working groups, summarized in Table \ref{table:WGoverview}, established ambitious goals for 2026 across multiple domains. The Calibration Statistics group aimed to publish the third Concordance paper and papers on multi-mission calibration (MCCal) and pileup effects using both Bayesian and neural network approaches. The Contamination working group targeted completion of a white paper draft by 2026 with plans to incorporate data from Einstein Probe and SVOM missions. The Non-thermal working group prioritized finishing the Crab Nebula paper, and the Coordinated Observations group focused on improving communications with new missions, while the Ground Calibration group worked on cross-checking raytrace packages against reference models and optics. The Detector and Background working group committed to updating background information on the wiki with links to current activities, and the Timing group requested updates to barycentric correction tools to support Suzaku, Hitomi, and XRISM observations. The Clusters of Galaxies working group recruited Ayşegül Tümer as deputy chair.

Paper development represented a critical component of IACHEC activities, with numerous manuscripts at various stages of completion. The third Concordance paper from the Calibration Statistics group was nearing submission to the Royal Astronomical Society, and the Crab cross-timing calibration paper had been published as Bachetti et al. (2026).

Alongside the contributed sessions, the IACHEC WGs held their meetings and this report summarizes the main results\footnote{The presentations are available at: \href{https://iachec.org/2026-seeblick-pelham-germany/}{\tt https://iachec.org/2026-seeblick-pelham-germany/}}, and comprises the reports from each of the IACHEC WGs: Calibration Uncertainties (\cref{s:calstat}), Detectors and Background (\cref{s:det_bkg}), Non-Thermal SNRs (\cref{s:nt_snr}), Thermal SNRs (\cref{s:t_snr}), Timing (\cref{s:timing}), and Isolated Neutron Stars and White Dwarfs (\cref{s:inswd}).

The IACHEC gratefully acknowledges sponsorship for the Society for Promotion of Space Science and the International Conference Support Program of Kanagawa Prefecture Japan, and from MPE and the \erosita\ mission.

\begin{table}
    \centering
    \begin{tabular}{| p{0.27\linewidth}|p{0.21\linewidth}|p{0.45\linewidth}|}
        \hline
         Working Group & WG Chair & Projects  \\
         \hline
         \hline
         Calibration Statistics & Vinay Kashyap & Quantifying response uncertainties; statistical methods; Concordance project \\
         \hline
         Clusters of Galaxies & Eric Miller \& Ayşegül Tümer & Multi Mission Study of selected targets from the HiFLUGCS sample\\
         \hline
         Contamination & Herman Marshall & Definition, measurement, and mitigation of molecular contaminant \\
         \hline
         Coordinated Observations & Karl Forster & Organization of yearly cross-calibration campaign of 3c273; investigation of potential cross-calibration candidate 1ES 0229+200; publication of 3c273 cross-calibration campaign\\
         \hline
         Detectors and Background & Catherine Grant & Forum for discussion of detector effects; background measurement \& modeling\\
         \hline
         Ground Calibration & Vadim Burwitz \& Takashi Okajima & comprehensive calibration baseline that will serve as a foundation for future missions\\
         \hline
         Heritage & Matteo Guianazzi & Curation of the IACHEC work; the IACHEC source database (\href{http://iachecdb.iaps.inaf.it}{\tt ISD})\\
         \hline
         Non-thermal SNR & Lorenzo Natalucci \& Craig Markwardt & Cross-calibration with the Crab and G21.5-0.3\\
         \hline
         Science Operations & Felix Fürst \& Hannah Earnshaw & Science operations processes and challenges, with a particular focus on aspects that affect calibration efforts\\
         \hline
         Thermal SNR & Paul Plucinsky  \& Hiromasa Suzuki & Cross-calibration with 1E 0102.2$-$7219 (E0102), N132D, and Cas~A; definition of standard models \\
         \hline
         Timing & Megumi Shidatsu \& Yukikatsu Terada & Summary of Timing performance and calibration status across missions; Systematic timing cross-calibration with Crab archive data; Planning timing cross-calibration campaign\\
         \hline
         White Dwarf and Isolated Neutron Stars & Vadim Burwitz & Cross-calibration with RX\,J1856.5$-$3754 and 1RXS\,J214303.7$+$065419 \\
         \hline
    \end{tabular}
    \caption{IACHEC Active Working Groups}
    \label{table:WGoverview}
\end{table}

\section{Working Group reports}

\subsection{Calibration Uncertainties: CalStats}\label{s:calstat}

This report on the Calibration and Statistics (CalStats) Working Group was presented by Vinay Kashyap from the Chandra X-ray Center at the Center for Astrophysics (CXC/CfA), who serves as chair, along with co-chair Ivan Valtchanov from the European Space Agency (ESA), at the 18th IACHEC meeting held at Seeblick Pelham on April 24, 2026. The CalStats Working Group serves as a forum for discussion of statistical, methodological, and algorithmic issues that affect the calibration of astronomical instruments, how calibration data are used in data analysis, and how analysis results are interpreted. The working group has grown to forty-five members, and interested parties can join by sending an email to the iachec-calstat google group at iachec-calstat+subscribe@cfa.harvard.edu, while those wishing to unsubscribe can email iachec-calstat+unsubscribe@cfa.harvard.edu. The working group maintains multiple communication and resource platforms including a main webpage at iachec.org/calibration-statistics with an extensive library, a wiki hosted at wikis.mit.edu that includes the tau matrix for Concordance analysis, background models and scripts, a Slack channel at iachec.slack.com under the hashtag calstats, and direct email contact at iachec-calstat@cfa.harvard.edu.

The working group reported on several significant activities and presentations during the meeting week. On Wednesday, April 22, Peter Boorman and Johannes Buchner delivered a tutorial on BXA (Bayesian X-ray Analysis) focused on statistical aspects of X-ray spectral analysis, with materials available at peterboorman.com/tutorial\_bxa. Also on Wednesday, Herman Marshall presented work on Concordance methodology applied to XMM-Newton and Chandra blazar cross-calibration, showing effective area corrections across different energy passbands for the XMM pn and MOS1 detectors using multiple spectral models. The working group also highlighted several innovative presentations throughout the meeting that showcased novel uses of statistical techniques, including Gabriel Matzeu's work on Monday, April 20, on cross-calibrating XMM pn and NuSTAR using hardness ratio trends in spectral model parameter estimates, Dimitrios Maniadakis's presentation on cross-calibration via energy-dependent pulse fractions of Vela X-1, Ayşegül Tümer's work on cross-calibration through flux correction trends over narrow passbands, Manan Agarwal's development of UltraSPEX with application to spatio-spectral modeling of Cassiopeia A presented on both Monday and Tuesday, and Megan Eckart's presentation on Tuesday, April 21, regarding estimation and use of XRISM Resolve gain and uncertainty.

In terms of updates to working group resources, the CalStats team maintains a library of useful papers curated by the working group leads, encouraging members to send pointers to any interesting papers, preprints, white papers, or calculations relevant to calibration statistics. A new addition to the library is a paper by M.\ Bonamente et al.\ published in 2026 (ApJ 1004, 11; arxiv 2604.02664) examining biases in the WStat method, which provides statistical analysis of three common methods of regression for Poisson data in the presence of Poisson background and finds that the non-parametric background method (wstat), which is popular for spectral data, is significantly biased especially in low-count and background-dominated regimes. Additional papers are anticipated soon, including work by X.\ Li et al.\ on cstat goodness-of-fit (submitted to Annals of Applied Statistics) and S.\ Das et al. on Concordance (submitted to RASTI). The wiki of background models has been updated with new entries for eROSITA, XRISM Resolve, and XRISM Xtend missions, and the wiki of the Concordance tau-matrix has received changes to entries for ROSAT, XMM pn, and NuSTAR.

Looking ahead to the road to IACHEC XIX, the working group outlined ambitious plans for the coming year. The group aims to complete several papers currently in progress, including work by J.\ Drake et al.\ on MCCal (Monte Carlo Calibration), R.\ Zimmerman et al.\ on pileup using a Bayesian approach, and J.\ Yang et al.\ on pileup using a neural network approach. The working group plans to develop and promote MCCal implementations across different mission analysis systems. A major focus will be organizing talks and tutorials on applications of artificial intelligence and machine learning to calibration problems, specifically addressing machine learning uncertainties distinguishing between aleatoric uncertainty (which represents randomness and data fluctuations) and epistemic uncertainty (which represents lack of knowledge and incomplete information), as well as exploring agentic AI, foundation models, and embeddings. The working group also emphasized the importance of synergies with other IACHEC working groups, particularly with the High-Resolution Spectroscopy Working Group regarding line locations for energy scale and gain calibration, the Ground Calibration Working Group on analysis calibration methods as referenced in Vadim Burwitz's presentation, and the Background Working Group on background models and their implementation.

\subsection{Clusters of Galaxies}\label{s:clusters}

The Galaxy Clusters Working Group studies the intracluster medium (ICM) of clusters as stable, broad-band X-ray standard candles for cross-calibration purposes. The group is chaired by Eric Miller and includes participants from almost all active and retired X-ray missions. The WG met for one session during the 18th IACHEC, featuring a single talk by Itsuki Aihara, who presented cross-calibration results comparing the XRISM Resolve and Xtend instruments with several other missions using nearby galaxy clusters. The comparison of derived ICM temperatures revealed potential calibration differences between instruments, akin to what has been previously reported (Nevalainen et al.\ 2010, Kettula et al.\ 2013, Schellenberger et al.\ 2015). Suzaku and Chandra measurements show excellent consistency with Resolve, agreeing within ±10\%. However, XMM-Newton measures temperatures 10--30\% lower than Resolve. A particularly notable finding was that Xtend temperatures tended to be higher than other CCD instruments. In a separate plenary talk, Jeremy Sanders presented work on improving eROSITA's in-flight calibration through source stacking techniques, with specific comparisons in Coma cluster regions using eROSITA, XMM-Newton EPIC-pn, and Chandra. The comparison showed systematic temperature differences between instruments. 

Both talks demonstrated the continued need for a coordinated cluster cross-calibration project. In that vein, the WG chair Eric Miller provided a status update on the Multi-Mission Study, an effort that began in 2017 to extend calibration comparisons across multiple missions and instruments using hot galaxy clusters selected from HiFLUGCS (Reiprich \& B\"ohringer 2002). The study employs specific selection criteria to ensure that systematic effects are minimized, including ICM $kT\geq6$ keV, $z\leq0.1$, more than 100,000 counts in the central 6 arcmin, and observations with cluster centers less than 3 arcmin off-axis. Contacts have been established across all participating observatories, and several missions have now supplied information about available observations that meet these criteria. A key priority for the Study is to provide standardized data products for the broader IACHEC Concordance effort (Section \ref{s:calstat}). To further facilitate progress along these lines, Ayşegül Tümer was appointed Deputy Chair of the WG.

\subsection{Detectors and Background}\label{s:det_bkg}

The Detectors and Background Working Group met at the 18th IACHEC meeting in April 2026, featuring five presentations addressing detector calibration and background characterization across multiple X-ray missions. The working group provides a forum for cross-mission discussion of detector-specific modeling and calibration issues, with emphasis on measuring and modeling instrument background in spatial, spectral, temporal, and orbital dimensions.

Michael Freyburg presented a detailed study on X-ray and particle background at the L2 Lagrange point, comparing eROSITA, XMM-Newton, and ACE data. This comprehensive investigation of flaring background phenomena is being prepared for publication and provides crucial insights into the radiation environment at L2, an increasingly popular orbital location for future X-ray missions.

Craig Markwardt presented the NICER SCORPEON Background Model, a physically-motivated, multi-component framework developed for NICER's complex low Earth orbit (LEO) environment. The modeling approach could potentially be adapted to other, possibly simpler, orbital configurations, suggesting broader applications across missions in various orbital environments.

Caroline Kilbourne discussed the in-orbit instrumental background of XRISM's Resolve instrument, also in LEO, highlighting its remarkably low background rates. While excellent for sensitive scientific observations, the low background makes correlation studies of the background itself difficult due to limited statistics. Kilbourne emphasized the importance of accurate atomic data in distinguishing instrumental background features from astrophysical signals in Resolve's high-resolution spectra.

Hiromasa Suzuki presented findings on flaring events observed by XRISM's Xtend instrument during night-Earth observations. These events occur primarily at high latitudes with 5-10 minute durations, possibly caused by precipitating electrons. The presentation included a valuable multi-mission background spectrum comparison providing context for understanding XRISM's background environment relative to other missions.

Megumi Shidatsu examined relative energy calibration accuracy between XRISM's Resolve and NuSTAR, finding that NuSTAR's energy offset from Resolve is comparable to systematic energy uncertainty. This highlights ongoing challenges in achieving precise energy calibration alignment across different detector technologies.

Possible topics to be discussed in future meetings include experience with and expectations for large focal planes containing hundreds to thousands of independent pixels (e.g. CMOS and X-IFU) and methods for their gain and spectral resolution calibration, and improvements in measuring and modeling the background at L1/L2, particularly in response to NewAthena's tight requirements on background knowledge.

\subsection{Non-Thermal SNRs}\label{s:nt_snr}

The Non-Thermal SNR Working Group met at the 18th IACHEC meeting in April 2026, led by Lorenzo Natalucci and Craig Markwardt. The group focused on cross-calibration using the Crab Nebula as a standard candle across X-ray and gamma-ray missions, particularly emphasizing energies above 10 keV.

James Rodi presented analysis of INTEGRAL Crab spectra, revealing significant calibration discrepancies. ISGRI and SPI instruments diverge above 350 keV, with different spectral slope measurements: SPI ($\Gamma$ = -2.32), ISGRI ($\Gamma$ = -2.58), and PICsIT ($\Gamma$ = -2.44) when fitting with GRBM models ($\Gamma$ = -1.99, Ec = 535 keV). Comparisons with COMPTEL data showed ISGRI decreasing toward COMPTEL values while PICsIT overlaps but measures ~1.7 times lower than SPI at ~1 MeV. This unclear MeV behavior makes calibration corrections non-obvious.

Craig Markwardt presented a novel flux transfer method using sky monitors and RXTE PCA as intermediate calibration standards. Since X-ray instruments with absolute calibration do not observe bright sources frequently, sky monitors can serve as relative transfer standards across time gaps. The technique uses NuSTAR stray light (2013-present), RXTE PCA (1996-2012), Swift BAT (2005-2018), and Fermi GBM (2008-present), all with ~1\% relative errors. An example using GBM to extend PCA flux calibration to the NuSTAR epoch showed ~4\% differences. However, combined sky monitor data normalized to Crab revealed clear systematic breaks between epochs.

The working group is conducting a comprehensive multi-mission Crab project following Kirsch et al. 2005, analyzing nearly simultaneous observations across 18 epochs spanning 2005-2020. The study emphasizes hard band energies ($>10$~keV) using power-law fits below 100 keV and broken power-law or GRBM models above 100 keV. An extensive contact list includes representatives from NuSTAR, INTEGRAL, Suzaku, HXMT, RXTE, Swift, NICER, AstroSat, XMM-Newton, Fermi GBM, Ninjasat, Einstein Probe, XPOSAT, and XRISM.

Recent progress on the Crab paper includes updated contact lists, revised NuSTAR spectra in the collaborative manuscript, and improved instrument sections, though INTEGRAL sections need revision and others require writing. The group decided to add NICER, XRISM, and XPOSAT data, create a single data repository, and hold regular teleconferences. The dataset now extends through 2025 with new epochs in 2022, 2024, and 2025, though XRISM, XPOSAT, and Ninjasat data remain to be incorporated. Remote meetings continue between IACHEC plenary meetings.

\subsection{Thermal SNRs}\label{s:t_snr}\label{s:n132d}
The Thermal Supernova Remnants (SNRs) WG meeting, lead by Paul Plucinsky (Center for Astrophysics $|$ Harvard-Smithsonian) and Hiromasa Suzuki (University of Miyazaki) was held on Tuesday, April 21, 2026. The focus of this WG is to obtain and maintain spectral models of several SNRs, which are useful for evaluating the effective area, detector gain, and the spectral redistribution function as these sources have line-rich spectra and are constant. The WG meeting covered several X-ray bright SNRs being used as calibration sources across multiple X-ray observatories including Chandra, XMM-Newton, Suzaku, eROSITA, XRISM, NuSTAR, and Swift. The agenda focused on five key SNRs.
For 1E~0102.2$-$7219 (E0102), significant updates were reported by Peter Ratzlaff (Center for Astrophysics $|$ Harvard-Smithsonian) and Paul Plucinsky regarding the inclusion of the \ion{Mg}{xi}~He$\alpha$ line complex normalization in the spectral fits for the first time. Previously, the spectral fits had only allowed the normalizations of the \ion{O}{vii}~He$\alpha$~line complex, the \ion{O}{viii}~Ly$\alpha$ line, the \ion{Ne}{ix}~He$\alpha$ line complex, and \ion{Ne}{X}~Ly$\alpha$ line to vary.  This change was motivated by the increasing absorption of the ACIS contamination layer which also resulted in the discontinuation of fitting the \ion{O}{vii}~He$\alpha$ normalization in 2021 and the \ion{O}{viii}~Ly$\alpha$ normalization in 2024. The investigation revealed that ACIS S3 (a back-illuminated CCD on Chandra) \ion{Mg}{xi}~He$\alpha$ line normalizations are 5--20\% higher than the IACHEC reference value.  Konrad Dennerl (MPE) showed that the eROSITA values and Eric Miller (MIT) showed that the Suzaku XIS values are also elevated with respect to the standard IACHEC value. Work is ongoing at the ESAC to investigate values from XMM-Newton's pn, MOS1, MOS2, and RGS instruments.
For N132D, Hiromasa Suzuki presented comparisons of XRISM/Resolve observations from two epochs (first in 2023 and second in 2025), finding that the data are mutually consistent. The parameters of the best-fit thermal spectral models were consistent with each other within 1--2$\sigma$ confidence levels. This confirms the stability of the microcalorimeter Resolve. 
The Resolve data clearly show different velocities and line broadening for intermediate-mass elements versus iron, with velocity dispersions around 450~km~s$^{-1}$ for silicon and sulfur versus 1670 ~km~s$^{-1}$ for iron. He also found that the latest IACHEC spectral model of N132D, with these broadening and redshift values applied, is mostly consistent with the Resolve data at around 6~keV, while reporting significant discrepancies in the 2--3~keV energy range.

Cas A was highlighted for groundbreaking XRISM observations that detected chlorine and potassium emission lines. Manan Agarwal's (University of Amsterdam) UltraSPEX analysis provided detailed measurements of energy shifts and line broadening in one-by-one arcminute regions, raising questions about how this high-resolution spectroscopic information could inform standard models for lower-resolution instruments. 
For Tycho's supernova remnant, Leila Godinaud (CEA Saclay) presented results from both Chandra and XRISM observations, discussing whether Tycho should be adopted more broadly as a calibration source by other missions, noting that Swift has already been using it. The presentation emphasized that while Tycho has advantages as a bright source with prominent silicon, sulfur, and iron emission lines and no temporal variation, there are spatial variations in line energy and broadening that need consideration. Models fitted to XRISM observations and existing Chandra parameter mapping provide good representations despite the spatial complexity. 
Finally, for Kepler's supernova remnant, Brian Grefenstette (Caltech) reported on NuSTAR's use of the iron K-line for calibration purposes.

Throughout the discussion, a recurring theme was the difficulty of translating the exceptional spectral resolution and detailed kinematic information from XRISM/Resolve into useful calibration standards for instruments with lower energy resolution. The challenge is to take advantage of the information provided by the high spectral resolution of the XRISM data, albeit with moderate spatial resolution, and information provided by the high spatial resolution of Chandra, albeit with moderate spectral resolution.

\subsection{Coordinated Observations}\label{s:coordobs}

Although the Coordinated Observations Working Group has not formally convened during the 18th IACHEC, several plenary talks by both longstanding and new IACHEC members addressed its scope, covering the coordinated cross-calibration status of NuSTAR, XRISM, XMM-Newton, and Chandra.
A. T"{u}mer presented the effective-area cross-calibration status derived from three years (2024--2026) of coordinated observations of the blazar 3C273. XRISM joined the campcampaign in 2024, and yearly effective-area cross-calibration results for NuSTAR, XRISM, XMM-Newton, and Chandra were shown for 2024--2025. In 2025 not all observatories overlapped, but observations were taken close enough in time, and the source showed no variability, that cross-comparison remained possible. The narrow-band flux ratio between Xtend and Resolve showed that their internal effective-area calibration has remained steady, with fluxes agreeing to within 5\% in the 2.3--10.0 keV band, rising to about 10\% in the softest (1.8--2.3 keV) and hardest (10.0--12.0 keV) bands. XMM-Newton measured the lowest flux with the highest photon index, while NuSTAR measured the highest flux, agreeing with Xtend and Resolve to within 5\%. XRISM's flux and photon index fell between the high-flux group (NuSTAR) and the low-flux group (XMM-Newton and Chandra/HETG). NuSTAR's FPMA and FPMB fluxes diverged from each other in 2025 but recovered agreement in the 2026 data. This 3C273 cross-calibration work is in preparation for submission to JATIS.
S. Inoue presented effective-area and quantum-efficiency cross-calibration results from a coordinated observation campaign of PKS 2155-304, with roughly 40 ks of overlap between XRISM and NuSTAR. As with the 3C273 work, XRISM and NuSTAR effective areas agreed to within 5\%, and the Xtend/Resolve flux ratio agreed to within 5\% in the 2.3--8.0 keV band, widening to 15\% at softer and harder energies. This cross-calibration work on G21.5 and PKS 2155-304 has been submitted to JATIS.

G. Matzeu presented a sample of coordinated XMM-Newton–NuSTAR observations (primarily Galactic sources), EPIC-pn in timing mode; NuSTAR FPMA/FPMB. Her X-1 analysis showed the majority of EPIC-pn slopes being softer with respect to NuSTAR FPMA/B and hard pn spectra tend do disagree most with FPMA. Flux difference of ~20\% was seen in most of the sources with some outliers. Slope difference of $\pm$0.1 for most of the targets with some pronounced outliers. Observations with harder $\Gamma$ were more likely to deviate from the above thresholds.

D. Maniadakis presented Vela X-1 analysis results from simultaneous NuSTAR and XMM-Newton EPIC-pn observations using Pulse Fraction spectrum as a cross-calibration diagnostic.

\subsection{Contamination}
This presentation on the Contamination Working Group Plan was delivered by Herman L. Marshall on April 23, 2026, at the 18th IACHEC meeting, outlining the working group's goals to update and compare contamination models across X-ray missions and generate a white paper on mitigation and analysis strategies. The working group reported varying contamination status across multiple missions: Chandra ACIS contamination continues to grow and requires model updates led by A. Bogdan and H. Marshall, XMM-Newton RGS contamination is no longer asymptotic but now increasing, and XRISM Xtend has only established an upper limit to contamination, indicating minimal issues. Marshall noted that white paper authors have been put on notice to begin finalizing the comprehensive documentation.

The presentation displayed detailed 2026 contamination results from multiple missions using different units, with the key conversion that optical depth $\tau = 0.5$ corresponds approximately to a thickness of 300 nanometers or a column density of 1.4$\times 10^{18}$ cm$^{2}$. Results were shown from Sanchez-Fernandez for RGS (showing growth from 2000-2015 reaching around 300 nanometers); Yoneyama for XRISM/XIS and Xtend (showing only upper limits through late 2025); Marshall for Chandra ACIS (demonstrating continued growth from 2000-2025); Rosen for XMM-Newton MOS1 and MOS2 (comparing 2022 and 2025 data); Zhang for Einstein Probe FXTA and FXTB (over 800 days since launch) reaching 0.5 or 0.35 optical depths, respectively; and Cheng/Ling for EP Wide Field Telescope showing the 0.4-0.6 keV rate decreased found in November 2025 was 63.5\% of that in February 2024, indicating significant contamination accumulation for one of the detectors.

The working group has established a concrete plan for the white paper being developed. The draft titled "Contamination on Detectors in X-ray Telescopes" with Marshall as lead author includes sections organized by mission status covering Chandra, Rossi XTE, XMM-Newton, Suzaku, AstroSat, Swift, NICER, NuSTAR, eROSITA, XRISM, IXPE, Einstein Probe, MAXI, HXMT, and future mission plans for Athena, SMILE/SXI, and eXTP. Marshall encouraged participants to edit the document during the meeting, with target completion by the next IACHEC plenary and intended publication in the Journal of Astronomical Telescopes, Instruments, and Systems (JATIS).
 
\subsection{Timing}\label{s:timing}

The Timing working group aims to provide a forum for sharing information and discussing timing calibration for X-ray missions both on the ground and in orbit. The group is currently coordinated by the chair, Megumi Shidatsu, and co-chair, Yukikatsu Terada. At this IACHEC workshop, the Timing working group session was held on 2026 April 22, the third day of the workshop. 

The working group provided comprehensive updates on ongoing and completed paper efforts in timing cross-calibration. A systematic timing cross-calibration paper led by Matteo Bachetti has been published in the Publications of the Astronomical Society of the Pacific as Bachetti et al. (2026). The paper describes a newly developed tool for timing cross-calibration and its methodology, together with its application to Crab observations from numerous missions, including Chandra, Einstein Probe, Fermi, Hitomi, Insight-HXMT, IXPE, MAXI, NICER, NinjaSat, NuSTAR, Suzaku, Swift, XMM-Newton, XRISM, and RXTE. In addition, papers on the XRISM timing calibration were published in the Journal of Astronomical Telescopes, Instruments, and Systems (JATIS) in 2025 (Terada et al. 2025; Shidatsu et al. 2025; Sawada et al. 2025). These papers include results from timing cross-calibration studies based on simultaneous observations of XRISM/Resolve with NICER and NuSTAR conducted in 2024 under the IACHEC collaboration.

In this session, the first presentation, given by Yugo Motogami, focused on verification of the timing accuracy of the XRISM Resolve instrument, based on the work in the JATIS paper series (2025) on XRISM timing performance and calibration. The talk reported that the new version of the XRISM calibration database (CALDB) achieved a significant improvement in the timing accuracy, reducing the standard deviation from 44.7 $\mu$s to 16.4 $\mu$s, confirming the effectiveness of the correction. The presentation also addressed transitions from the GPS-synchronized mode to the GPS-unlocked mode that occurred in March 2026. Although two short GPS-unlocked intervals occurred on March 24, 2026, the system successfully returned to GPS synchronization afterward, and no timing issues were detected outside those intervals.

Matteo Bachetti delivered a remote presentation introducing the paper Bachetti et al. (2026), with both the Time-of-arrival (TOA) calculation code\footnote{github.com/matteobachetti/TOAextractor} and deadtime correction code\footnote{github.com/matteobachetti/pulse\_deadtime\_fix} now publicly accessible on GitHub. The next steps for this work include expanding the tool's application beyond just the Crab pulsar to include more astronomical sources and conducting a deeper analysis of the results for each instrument. 

At the end of the session, we reviewed the points of contact for each mission and discussed future observation plans for timing calibration purposes. The XPoSat mission joined the collaboration, and the information was added to the Timing Working Group wiki\footnote{\url{wikis.mit.edu/confluence/display/iachec/Timing}}. 
We also revisited the two action items that were raised at the 2025 IACHEC workshop: updating the solar system ephemeris used in the barycentric correction tool {\tt barycen} in HEAsoft from DE-200 to DE-430, and transitioning the Crab pulsar ephemeris from the Jodrell Bank ephemeris to the Fermi ephemeris. The former has already been requested to the developers in HEASARC at NASA/GSFC, although it has not yet been implemented. The latter remains a task for future work and would be valuable for timing analyzes of other pulsars.

\subsection{Isolated Neutron Stars and White Dwarfs}\label{s:inswd}

This WG aims to improve the cross-calibration of X-ray telescopes in the low energy range ($<$ 1.5 keV) by using spectra of Isolated Neutron Stars (INSs) and White Dwarfs (WDs). These objects should not display time-dependent variation and should have physically well-modeled spectra that can be used as spectral standard candles at low energies. Over the years, a set of white dwarfs (GD153, Sirius B, and HZ43) with spectra that can be described by physical white dwarf models, and the isolated neutron star RX\,J1856.5$-$3754 (RX\,J1856) with a spectrum that can best be described by a single black-body model, have met the characteristics of a standard candle. The three WDs and RX\,J1856  were used to improve the calibration of the low energy end of the \chandra\ LETGS (LETG+HRC-S) and provide a cross-calibration with \rosat\ and recently also with \nicer\  and \erosita, and the \EP\ Follow-up X-ray Telescope. In the context of \erosita\ a new INS is being studied and used for calibration: 1RXS\,J214303.7$+$065419 (RX\,J2143). 

Surodeep Sheth (AIP) gave a very detailed overview RX\,J1856 as a cross-calibration standard candle. Specifically comparing eROSITA calibration observations of RXJ1856 with those obtained with other X-ray observatories (Chandra LETG, XMM-Newton, ASTROSAT).

Herman Marshall (MIT) showed an update of ACIS CCD contamination monitoring using observations of RX\,J1856.

In summary, the INS RX\,J1856.5$-$3754 and white dwarf HZ43 are being observed and used by many X-ray observatories for calibration purposes to monitor the status of their low energy calibration. 

\subsection{Ground Calibration "GCal" Working Group}\label{s:gcal}

This Ground Calibration Working Group summary was presented by Vadim Burwitz (MPE) and Takashi Okajima (GSFC) on behalf of the working group members. The substantial progress and ongoing activities of the Ground Calibration Working Group were presented. The working group has been making significant advances in several topics designed to improve consistency, reliability, and standardization of ground calibration procedures across different X-ray missions and test facilities worldwide.

In the area of X-ray test facilities, the working group has made notable progress in cross-calibration using a reference optic, successfully identifying a Nickel electroformed optic as the standard reference for cross-calibration across different test facilities. A paper on segmented optics testing has been proposed. Further cross-calibration work is actively ongoing for several major future missions, including ATHENA (the Advanced Telescope for High-Energy Astrophysics), Einstein Probe (EP), and AXIS with its Next Generation X-ray Optics (NGXO). For ray-tracing packages, the working group has identified the need to cross-check different software tools against a reference optic model, though this work has not yet been initiated and remains on the agenda for future activities. In the area of analysis methods, the working group is verifying Half-Energy Width (HEW) computation techniques using a master data set, and has proposed a paper on this topic that could reportedly advance swiftly, suggesting that the groundwork has been well prepared. Finally, the working group is developing guidance documentation for new missions to help them develop comprehensive and effective ground calibration plans, ensuring that future X-ray observatories benefit from the collective experience and best practices established by previous missions and the calibration community.

\subsection{Science Operations Working Group}

The Science Operations working group is concerned with discussing common challenges and new concepts related to operating X-ray telescopes in space. A particular focus is the impact of operational constraints (e.g., thermal constraints, changes in orbital altitude) on the calibration of the instruments. The working group is currently co-led by Felix Meeker-F\"urst (ESA) and Hannah Earnshaw (Caltech) and met at the 18th IACHEC meeting for the first time in person. 

Before the meeting, the working group had met once online, where presentations were given from the operations teams of NuSTAR and Swift, as well as a status update on the NASA ACROSS facility. At the IACHEC, XMM-Newton (Felix Meeker-F\"urst), NuSTAR (Hannah Earnshaw), and XRISM (Yukikatsu Terada) presented their current operations status. For NICER, Kenji Hamaguchi described the development of a ToO webpage reporting system.

The working group leaders would like to expand participation for the next annual meeting, encouraging all other missions to contribute presentations about their operations experiences and practices. The working group plans to hold two future online meetings during the year to address specific operational themes. A teleconference scheduled for around October 2026 will focus on rejuvenation and modernization activities, exploring strategies and best practices for keeping aging missions up-to-date and operationally viable. A second teleconference planned for around January 2027 will concentrate on long-term planning, examining best practices for observation scheduling and discussing the potential use of artificial intelligence in future mission operations, addressing the question of how operational planning will evolve in the coming years. 

Beyond these scheduled teleconferences, the working group identified several additional topics of interest for future discussions. These include capturing and sharing lessons learned from previous and current missions to inform the planning of new missions and improvements in the operation of current missions. The working group also recognized the need for better communication and coordination between missions regarding short-term schedule changes, particularly for ToO observations where rapid coordination among multiple observatories can significantly enhance scientific return. Anyone interested in joining the SciOps Working Group and participating in these discussions is encouraged to contact Hannah Earnshaw directly via email at hpearn@caltech.edu to be added to the distribution list for future meetings and teleconferences.

\section*{References\footnote{see \href{https://iachec.org/papers/}{\tt https://iachec.org/papers/} for a complete list of IACHEC papers}}

\noindent
Audard, M., et al.\ 2024, \href{https://doi.org/10.1093/pasj/psae080}{PASJ, 76, 1186}\\
\noindent
Bachetti, M., et al.\ 2026, \href{https://doi.org/10.1088/1538-3873/ae2b36}{PASP, 138, 014501} \\
\noindent
Beuermann, K., Burwitz, V., Rauch, T. 2006, 
\href{https://doi.org/10.1051/0004-6361:20065478}{A\&A 458, 541}\\
\noindent
Chen, Y., et al.\ 2019, \href{https://doi.org/10.1080/01621459.2018.1528978}{J.Am.Stat.Assoc., 114:527, 1018}\\
\noindent
Jourdain, E.\ \& Roques, J. P.\ 2020, \href{https://doi.org/10.3847/1538-4357/aba8a4}{ApJ, 899, 131}\\
\noindent
Kaastra, J., et al.\ 2009, \href{https://doi.org/10.1051/0004-6361/20077892}{A\&A, 497, 291}\\
\noindent
Kettula, K., Nevalainen, J., \& Miller, E.D.\ 2013, \href{https://doi.org/10.1051/0004-6361/201220408}{A\&A, 552, A47}\\
\noindent
Kuiper, L., et al.\ 2001, \href{https://doi.org/10.1051/0004-6361:20011256}{A\&A, 378, 918}\\
\noindent
Marshall, H., et al.\ 2021, \href{https://doi.org/10.3847/1538-3881/ac230a}{AJ, 162, 254} \\
\noindent
Marshall, H.\ 2024, \href{https://doi.org/10.3847/1538-4357/ad0897}{ApJ, 964, 88} \\
\noindent
Nevalainen, J., David, L., \& Guainazzi, M.\ 2010, \href{https://doi.org/10.1051/0004-6361/201015176}{A\&A, 523, A22}\\
\noindent
Osborne, C.\ 1991, \href{https://doi.org/10.2307/1403690}{International Statistical Review, 59, 309}\\
\noindent
Reiprich, T.H. \& B\"ohringer, H.\ 2002, \href{https://doi.org/10.1086/338753}{ApJ, 567, 716}\\
\noindent
Schellenberger, G., et al.\ 2015, \href{https://doi.org/10.1051/0004-6361/201424085}{A\&A, 575, A30}\\
\noindent
Sawada, M., et al.\ 2025, \href{https://doi.org/10.1117/1.JATIS.11.4.042011}{JATIS, 11, 042011}\\
\noindent
Shidatsu, M., et al.\ 2025, \href{https://doi.org/10.1117/1.JATIS.11.4.042012}{JATIS, 11, 042012}\\
\noindent
Suzuki, H., et al.\ 2020, \href{https://doi.org/10.3847/1538-4357/aba524}{ApJ, 900, 39}\\
\noindent
Terada, Y., et al.\ 2025, \href{https://doi.org/10.1117/1.JATIS.11.4.042007}{JATIS, 11, 042007}\\
\noindent
Tsujimoto, M., et al.\ 2011, \href{https://doi.org/10.1051/0004-6361/201015597}{A\&A, 525, 25} \\
\noindent
Wilson-Hodge, C., et al.\ 2011, \href{https://doi.org/10.1088/2041-8205/727/2/L40}{ApJ, 727, 40}\\
\\

\end{document}